\documentclass[aps,twocolumn]{revtex4}
\usepackage{epsfig}
\usepackage{graphicx}
\usepackage{amsmath}
\usepackage{booktabs}

\begin{document}
\title{Towards the understanding of fully-heavy tetraquark states from various models}
\author{Chengrong Deng$^{a}{\footnote{crdeng@swu.edu.cn}}$,
        Hong Chen$^a{\footnote{chenh@swu.edu.cn}}$,
       and Jialun Ping$^b{\footnote{jlping@njnu.edu.cn}}$}

\affiliation{$^a$Department of Physics, Southwest University, Chongqing 400715, China}
\affiliation{$^b$Department of Physics, Nanjing Normal University, Nanjing 210097, China}

\begin{abstract}
We use a color-magnetic interaction model (CMIM), a traditional constituent quark model (CQM) and a multiquark color flux-tube model (MCFTM) to systematically investigate the properties of the states $[Q_1Q_2][\bar{Q}_3\bar{Q}_4]$ ($Q=c,b$). The dynamical investigation indicates that the CMIM can not completely absorb QCD dynamical effects through the effective constituent quark mass and overestimates the color-magnetic interaction in the states under the assumption of the same spatial configurations. The Coulomb interaction plays a critical role in the dynamical model calculations on the heavy hadrons, which induces the fact that none of bound states $[Q_1Q_2][\bar{Q}_3\bar{Q}_4]$ can be found in
the dynamical models. The color configuration $\left[[Q_1Q_2]_{\mathbf{6}_c}[\bar{Q}_3\bar{Q}_4]_{\bar{\mathbf{6}}_c}\right]_{\mathbf{1}}$ should be taken seriously in the ground states due to the strong Coulomb attraction between the $[Q_1Q_2]_{\mathbf{6}_c}$ and $[\bar{Q}_3\bar{Q}_4]_{\bar{\mathbf{6}}_c}$. The color configuration $\left[[Q_1Q_2]_{\bar{\mathbf{3}}_c}[\bar{Q}_2\bar{Q}_4]_{\mathbf{3}_c}\right]_{\mathbf{1}}$ is absolutely dominant in the excited states because of the strong Coulomb attraction within the $[Q_1Q_2]_{\bar{\mathbf{3}}_c}$ and $[\bar{Q}_2\bar{Q}_4]_{\mathbf{3}_c}$. The $J/\Psi$-pair resonances recently observed by LHCb are difficult to be accommodated in the CMIM. The broad structure ranging from 6.2 to 6.8 GeV can be described as the ground tetraquark state $[cc][\bar{c}\bar{c}]$ in the various dynamical models. The narrow structure $X(6900)$ can be identified as the excited state $[cc][\bar{c}\bar{c}]$ with $L=1$ ($L=2$) in the CQM (MCFTM).
\end{abstract}

\maketitle

\section{Introduction}

The dynamics in the fully-heavy tetraquark states is very simple, which only includes perturbative QCD one gluon exchange (OGE) interaction and quark confinement potential. They can provide a unique environment to examine the non-relativistic quark model with QCD effective potentials if they do exist. The question of whether there exist such states has been debated for more than forty years~\cite{iwasaki}, which received much attention from the different theoretical frameworks, such as the non-relativistic quark models~\cite{nrqm,lloyd,gjwang,xhzhong}, the color-magnetic interaction model~\cite{ brac,berezhnoy,karliner,cmwj}, the QCD sum rules~\cite{qcdsm}, the Bethe-Salpeter equation~\cite{bseq}, MIT bag model~\cite{mit}, the lattice QCD~\cite{4Qlqcd} et al. The conclusions were model dependent. Taking the state $bb\bar{b}\bar{b}$ as an example, it
can exist as a stable state against strong interaction in the spin-spin interaction model~\cite{spin-spin} while it is not stable in the string model~\cite{string}. Due to these controversial issues, relevant experimental studies are significant to understand their properties.

On the experimental side, the ATLAS, CMS and LHCb collaborations have measured the cross section for double charmonium production~\cite{exp4c}. In 2018, the LHCb collaboration investigated the $\Upsilon \mu^+\mu^-$ invariant-mass distribution to search for a possible fully-heavy tetraquark state $bb\bar{b}\bar{b}$, and observed no significant excess~\cite{exp4b}. Very recently, the LHCb Collaboration researched the invariant mass spectrum of $J/\Psi$ pairs using proton-proton collision data at centre-of-mass energies
of $\sqrt{s}=7$, 8 and 13 TeV~\cite{exp4c2}. They found a broad structure ranging from 6.2 to 6.8 GeV and a narrow structure around 6.9 GeV. The narrow structure, denoted as $X(6900)$, is assumed as a resonance with the Breit-Wigner lineshape. The structures are the first to be made up of four heavy quarks of the same type, which provides an extreme and yet theoretically fairly simple case to explore the strong interaction and to test models that can be used to explain the nature of ordinary hadrons.

The structures have revitalized the investigations of multiquark resonances made of heavy quarks and heavy antiquarks. Many theoretical investigations have been devoted to interpret
the properties of the structures~\cite{theory,segovia}. However, their properties and spin-parity quantum numbers are not completely clear so far. It is therefore necessary to carry out a dynamical investigation on the fully-heavy tetraquark states from various theoretical frameworks, which is propitious to identify the properties of the structures and to broaden our horizons of the fully-heavy exotic hadrons.

In this work, we prepare to make a systematical research on the states $[Q_1Q_2][\bar{Q}_3\bar{Q}_4]$ from the perspective of the phenomenological models, including the color-magnetic interaction model, traditional constituent quark model and multiquark color flux-tube model. The color-magnetic interaction models have various versions~\cite{ppnp}, the model with reference mass scale is employed here~\cite{referencemass}. The traditional constituent quark model includes the OGE interaction and two-body confinement potential proportional to color charge. The multiquark color flux-tube model based on the lattice QCD color flux-tube picture and the traditional quark model has been developed, which contains a multibody confinement potential instead of two-body one. The model was recently applied to systematically investigate the states $[cs][\bar{c}\bar{s}]$~\cite{cscs}. Furthermore, the conclusions of other phenomenological models are involved to make a comprehensive understanding on the fully-heavy tetraquark states.

This paper is organized as follows. After the introduction section, the descriptions of three models are given in Sec. II. The wavefunction of the states $[Q_1Q_2][\bar{Q}_3\bar{Q}_4]$ is shown in Sec. III. The numerical results and discussions are presented in Sec. IV. A brief summary is listed in the last section.

\section{Three Models}
\subsection{Color-magnetic interaction model (CMIM)}

The original form of the color-magnetic interaction between the particles $i$ and $j$ can be expressed as follows,
\begin{eqnarray}
V_{ij}^{cm} & = &-{\frac{\pi\alpha_{s}\delta(\mathbf{r}_{ij})\mathbf{\lambda}^c_{i}\cdot\mathbf{\lambda}_{j}^c\mathbf{\sigma}_{i}\cdot\mathbf{\sigma}_{j}}{6m_im_j}},
\end{eqnarray}
it is the spin-dependent part of the OGE interaction $V_{ij}^{oge}$~\cite{root},
\begin{eqnarray}
V_{ij}^{oge} & = & {\frac{\alpha_{s}}{4}}\mathbf{\lambda}^c_{i}\cdot\mathbf{\lambda}_{j}^c\left({\frac{1}{r_{ij}}}-
{\frac{2\pi\delta(\mathbf{r}_{ij})\mathbf{\sigma}_{i}\cdot
\mathbf{\sigma}_{j}}{3m_im_j}}\right),
\end{eqnarray}
$m_i$  and $\mathbf{r}_i$ are the effective mass and position of the particle $i$, respectively.
$\mathbf{r}_{ij}=\mathbf{r}_i-\mathbf{r}_j$ and $r_{ij}=|\mathbf{r}_{i}-\mathbf{r}_{j}|$. $\mathbf{\lambda}^c$ and $\sigma$ represent the Gell-Mann matrices and the Pauli matrices, respectively. $\alpha_s$ is a running quark-gluon coupling constant in the perturbative QCD. More details will be given in the following introduction of the constituent quark model.

With the exception of the spin-color factor $\mathbf{\lambda}^c_{i}\cdot\mathbf{\lambda}^c_{j}\mathbf{\sigma}_{i}\cdot\mathbf{\sigma}_{j}$, the other part of the color-magnetic interaction can be denoted as
\begin{eqnarray}
C_{ij} & = &{\frac{\pi\alpha_{s}\delta(\mathbf{r}_{ij})}{6m_im_j}}, \nonumber
\end{eqnarray}
It incorporates the effects from the spatial configuration and effective quark masses and describes the effective coupling constant between the particles $i$ and $j$. For $n$-body ground states, the total color-magnetic interaction can be written as
\begin{eqnarray}
H_{cm}^n &=& -\sum_{i<j}^{n}C_{ij}\mathbf{\lambda}^c_{i}\cdot\mathbf{\lambda}^c_{j}\mathbf{\sigma}_{i}\cdot\mathbf{\sigma}_{j},
\end{eqnarray}
It leads to the mass splitting among different color-spin configurations. In the conventional mesons and baryons, the color factors are frozen because they are constant,
$\langle\mathbf{\lambda}^c_i\cdot\mathbf{\lambda}^c_j\rangle=-\frac{8}{3}$ for baryons and $\langle\mathbf{\lambda}^c_i\cdot\mathbf{\lambda}^c_j\rangle=-\frac{16}{3}$ for mesons.
The calculation of color-magnetic interaction therefore reduces to the simple algebra of the spin-spin operator $\mathbf{\sigma}_{i}\cdot\mathbf{\sigma}_{j}$. In the case of the tetraquark state with diquark-antiquark configuration, the values $\langle\mathbf{\sigma}_{i}\cdot\mathbf{\sigma}_{j}\rangle$ and $\langle\mathbf{\lambda}^c_i\cdot\mathbf{\lambda}^c_j\rangle$ can be calculated according to the symmetry properties of the tetraquark wave function~\cite{symmetry}. In addition, they can also be calculated according to the definition of spin and color operators with our computer programs. A multiquark state with given total quantum numbers in general consists of several channels with different intermediate quantum numbers. Using the values $\langle\mathbf{\sigma}_{i}\cdot\mathbf{\sigma}_{j}\rangle$, $\langle\mathbf{\lambda}^c_i\cdot\mathbf{\lambda}^c_j\rangle$ and $C_{ij}$, the CMI matrices can be obtained~\cite{referencemass}. Finally, the color-magnetic interaction energies of the multiquark states can be achieved after we diagonalize the numerical CMI matrices.

The effective quark masses should also be involved in the CMIM Hamiltonian, which is assumed to be able to absorb other various QCD dynamic effects, such as kinetic energy, Coulomb interaction and confinement potential. Therefore, the mass formula of $n$-body ground states in the CMIM reads,
\begin{eqnarray}
M=\sum_{i=1}^n  m_i+\langle H_{cm}^n\rangle \label{mass}
\end{eqnarray}

In principle, the values of $m_i$ and $C_{ij}$ should be different in the various hadron environment. However, it is difficult to exactly obtain the effect from the spatial configuration because of no knowing the spatial wave function. Therefore, the CMIM assumes that two pairs of interactional particles with the same quark content share the same
size in the generalization from conventional hadrons to multiquark states. For the simplicity and model universality, the values of $m_i$ and $C_{ij}$ are usually extracted from
the masses of conventional hadrons and then extended to multiquark states~\cite{ppnp}. This mechanism has been applied to investigate the properties of the well-known H-particle
and heavy pentaquark state $qqqq\bar{Q}$~\cite{h-particle,pentaquark}. Recently, the mechanism was also widely utilized to study the natures of some new hadrons~\cite{highmass}.

The mass formula generally overestimate the mass of conventional hadrons~\cite{ppnp}, especially the mesons. In the case of various multiquark states, the obtained masses with
this mass formula are also the largest values~\cite{highmass}. The reason of the overestimation on the mass is probably from the fact that the dynamical effects cannot be simply absorbed into the effective quark masses. In order to reduce the uncertainties and obtain more appropriate estimations, an alternative mass formula has been developed to avoid generally overestimated masses~\cite{referencemass},
\begin{eqnarray}
M=M_{ref}-\langle H_{cm}^n\rangle_{ref}+\langle H_{cm}^n\rangle
\end{eqnarray}
$M_{ref}$ and $\langle H_{cm}^n\rangle_{ref}$ are the physical mass of the reference system and its corresponding color-magnetic interaction energy, respectively. A multiquark
(tetraquark) state is generally related to a reference hadron-hadron (meson-meson) system whose quark content and quantum numbers are the same as those of the considered multiquark state. This mass formula can evade the problem of using extracted quark masses from conventional hadrons in the multiquark states~\cite{evade}. Meanwhile, it can phenomenologically compensate the part of missed attraction between quark components in the multiquark states.

In the present work, we focus on the CMIM results of the states $[Q_1Q_2][\bar{Q}_3\bar{Q}_4]$ obtained from the reference mass formula. The parameters $C_{ij}$ related to ground heavy-meson states are taken from Ref.~\cite{ppnp}, some of which are used here are listed in Table \ref{cij}.

\begin{table}[ht]
\caption{Parameters $C_{ij}$ for the ground heavy-meson states, unit in MeV.} \label{cij}
\begin{tabular}{ccccccccccc}
\toprule[0.8pt] \noalign{\smallskip}
~~$C_{ij}$~~&~~~$C_{c\bar{q}}$~~~&~~$C_{c\bar{s}}$~~&~~~$C_{\bar{c}c}$~~~&~~$C_{\bar{b}q}$~~&~~~$C_{\bar{b}s}$~~~&~~$C_{\bar{b}b}$~~&~~~$C_{\bar{b}c}$~~~  \\
 Value   &        $6.6$         &      $6.7$     &      $5.3$           &      $2.1$     &      $2.3$           &      $2.9$     &      $3.3$      \\
\toprule[0.8pt] \noalign{\smallskip}
\end{tabular}
\end{table}

According to the alternative mass formula, one can define the binding energy of a tetraquark state as
\begin{eqnarray}
\Delta E=M-M_{ref}=\langle H_{cm}^n\rangle-\langle H_{cm}^n\rangle_{ref}
\end{eqnarray}
to identify whether or not the state is stable against strong interaction. If $\Delta E\geq0$, the state can fall apart into the two mesons through quark rearrangement. If $\Delta E<0$, the strong decay into the two mesons is forbidden and therefore the decay must be weak or electromagnetic interaction. In fact, such an estimation method had been applied to search for stable multiquark states  many years ago~\cite{origin}.

\subsection{Constituent quark model (CQM)}

Constituent quark model is formulated under the assumption that hadrons are color singlet nonrelativistic bound states of constituent quarks with effective masses and interactions. One expects that the model dynamics is governed by QCD. The perturbative effect of QCD is well described by the OGE interaction, which is a standard color Fermi-Breit interaction given by the Lagrangian
\begin{eqnarray}
L&=&i\sqrt{4\pi}\alpha_s\bar{\psi}\gamma_{\mu}G^{\mu}\lambda^c\psi
\end{eqnarray}
where $G^{\mu}$ is the gluon field. From the non-relativistic reduction of the OGE diagram in QCD for point-like quarks one gets
\begin{eqnarray}
V_{ij}^{oge} & = & {\frac{\alpha_{s}}{4}}\mathbf{\lambda}^c_{i}\cdot\mathbf{\lambda}_{j}^c\left({\frac{1}{r_{ij}}}-
{\frac{2\pi\delta(\mathbf{r}_{ij})\mathbf{\sigma}_{i}\cdot
\mathbf{\sigma}_{j}}{3m_im_j}}\right),
\end{eqnarray}
Dirac $\delta(\mathbf{r}_{ij})$ function comes out in the deduction of the interaction between point-like quarks, when not treated perturbatively, which leads to collapse~\cite{collapse}. Therefore, the $\delta(\mathbf{r}_{ij})$ function can be regularized in the form
\begin{equation}
\delta(\mathbf{r}_{ij})\rightarrow\frac{1}{4\pi r_{ij}r_0^2(\mu_{ij})}e^{-r_{ij}/r_0(\mu_{ij})},
\end{equation}
where $r_0(\mu_{ij})=\hat{r}_0/\mu_{ij}$, $\hat{r}_0$ is an adjustable model parameter. This regularization is justified based on the finite size of the constituent quarks and should be therefore flavor dependent~\cite{flavor-dependent}.

The quark-gluon coupling constant $\alpha_s$ in the perturbative QCD reads~\cite{alphas},
\begin{equation}
\alpha_s(\mu^2)=\frac{1}{\beta_0\ln\frac{\mu^2}{\Lambda^2}},
\end{equation}
In the present work, we use an effective scale-dependent form given by,
\begin{equation}
\alpha_s(\mu^2_{ij})=\frac{\alpha_0}{\ln\frac{\mu_{ij}^2}{\Lambda_0^2}},
\end{equation}
where $\mu_{ij}$ is the reduced mass of two interacting particles, namely $\mu_{ij}=\frac{m_im_j}{m_i+m_j}$. $\Lambda_0$ and $\alpha_0$ are adjustable model parameters.

With the exception of the OGE interaction, CQM model imitating QCD should also incorporate nonperturbative effect, color confinement, which takes into account the fact that the only observed hadrons are color singlets. Color confinement plays an essential role in the low energy hadron physics. However, it is still impossible for us to derive color confinement analytically from its QCD Lagrangian so far. In the CQM, it can be phenomenologically described as the sum of two-body interactions proportional to the color charges and $r_{ij}^2$~\cite{isgur-karl},
\begin{eqnarray}
V^{con}&=& -a_c\sum_{i<j}^n\mathbf{\lambda}^c_{i}\cdot\mathbf{\lambda}^c_{j}r^2_{ij}
\end{eqnarray}
where $a_c$ is an adjustable model parameter.

To sum up, the completely Hamiltonian of the CQM for the heavy mesons and fully-heavy tetraquark states can be presented as
\begin{eqnarray}
H_n =\sum_{i=1}^n \left(m_i+\frac{\mathbf{p}_i^2}{2m_i}\right)-T_{c}+\sum_{i<j}^n V_{ij}^{oge}+V^{con}.
\end{eqnarray}
$T_{c}$ is the center-of-mass kinetic energy of the states and should be deducted; $\mathbf{p}_i$ is the momentum of the ith quark or antiquark.

The model can automatically prevent overall color singlet multiquark states disintegrating into several color subsystems by means of color confinement with an appropriate $SU_c(3)$ Casimir constant~\cite{drawback}. The model also allows a multiquark system disintegrating into color-singlet clusters, and it leads to interacting potentials within mesonlike $q\bar{q}$ and baryonlike $qqq$ subsystems in accord with the empirically known potentials~\cite{drawback}. However, the model is known to be flawed phenomenologically because it leads to power law van der Waals forces between color-singlet hadrons. In addition, it also leads to anticonfinement for symmetrical color structure in the multiquark system~\cite{anticonfinement}.

In order to avoid the misjudgement of the behavior of model dynamics due to inaccurate numerical results, a high precision computational method is therefore indispensable. The Gaussian expansion method (GEM)~\cite{GEM}, which has been proven to be a rather powerful numerical method to solve few-body problem in nuclear physics, is therefore widely used to study multibody systems. According to the GEM, the relative motion wave function between the quark and antiquark in the heavy mesons can be written as,
\begin{eqnarray}
\phi^G_{lm}(\mathbf{r})=\sum_{n=1}^{n_{max}}c_{n}N_{nl}r^{l}e^{-\nu_{n}r^2}Y_{lm}(\hat{\mathbf{r}})
\end{eqnarray}
Gaussian size parameters are taken as geometric progression
\begin{eqnarray}
\nu_{n}=\frac{1}{r^2_n}, &r_n=r_1a^{n-1},
&a=\left(\frac{r_{n_{max}}}{r_1}\right)^{\frac{1}{n_{max}-1}}
\end{eqnarray}
$N_{nl}$ is normalized coefficient and $c_n$ is a variation coefficient determined by the model dynamics. More details about the GEM can be found in Ref.~\cite{GEM}. With $r_1=0.2$ fm, $r_{n_{max}}=2.0$ fm and $n_{max}=7$, the converged numerical results can be achieved in the present work.

The mass of $ud$-quark is taken to be one third of that of nucleon, other adjustable model parameters in Table \ref{para} can be determined by solving twobody Schr\"{o}dinger equation with the trial wave function Eq.(13) to fit the ground heavy meson states, which is listed in Table \ref{spectra}. At the same time, we also give the average values of various parts of the CQM Hamiltonian. $\langle E_k\rangle$, $\langle V^{con}\rangle$, $\langle V^{cm}\rangle$ and $\langle V^{clb}\rangle$ represent the average values of kinetic energy, confinement potential, color-magnetic interaction and Coulomb interaction, respectively. It can been found from Table \ref{spectra} that the Coulomb interaction provides an extremely strong short-range attraction, which is the main reason why a quark and an antiquark can form a bound state in the CQM.

In the CMIM, the matrix elements $\langle\sigma_i\cdot\sigma_j\rangle=-3$ and $1$ for spin $S=0$ and spin $S=1$, respectively. Under the assumption of the same spatial configuration, the color-magnetic interaction gives an energy ratio $3:1$ between spin $S=0$ and $S=1$ mesons with the same quark content in the CMIM, such as $D$ and $D^*$.
In the dynamical calculation of the CQM, the same flavor mesons with different spin do not share the same spatial configuration because of their different dynamics, see the average
distance $\langle r^2\rangle^\frac{1}{2}$ in Table \ref{spectra}. The ratio in the dynamical calculation is not strict $3:1$ because of the same reason, which is between $3:1$ and $4:1$. It is therefore approximately reasonable to describe the mass splitting of mesons by the color magnetic interaction in the CMIM.

\begin{table}[ht]
\caption{Model parameters, quark mass and $\Lambda_0$ unit in MeV, $a_c$ unit in MeV$\cdot$fm$^{-2}$, $r_0$ unit in MeV$\cdot$fm and $\alpha_0$ is dimensionless.}\label{para}
\begin{tabular}{ccccccccccc}
\toprule[0.8pt] \noalign{\smallskip}
Para.   &~$m_{u,d}$~ & ~~$m_{s}$~~ & ~~$m_c$~~  &  ~~$m_b$~~   &  ~~$a_c$~~  &  ~~$\alpha_0$~~  & ~~$\Lambda_0$~~   & ~~$r_0$~~  \\
Valu.   &     313    &     494     &   1664     &    5006      &    $-150$   &       4.25       &         40.85     &   119.3    \\
\toprule[0.8pt] \noalign{\smallskip}
\end{tabular}
\caption{Ground state heavy-meson spectra and the average values of various parts of the Hamiltonian in MeV and the average distance in fm.} \label{spectra}
\begin{tabular}{ccccccccccccc}
\toprule[0.8pt] \noalign{\smallskip}
States         & ~~PDG~~ &~~$E_2$~~&~~$\langle E_k\rangle$~~&$\langle V^{con}\rangle$&$\langle V^{cm}\rangle$&$\langle V^{clb}\rangle$&~$\langle r^2\rangle^\frac{1}{2}$~\\
$D^{\pm}$      &  1869   &   $1886$  &    737    &       200     &   $-92$      &    $-937$     &  0.50   \\
$D^*$          &  2007   &   $2000$  &    633    &       226     &   $ 27$      &    $-862$     &  0.53   \\
$D_s^{\pm}$    &  1969   &   $1982$  &    693    &       151     &   $-105$     &    $-914$     &  0.43   \\
$D_s^*$        &  2112   &   $2109$  &    560    &       179     &   $ 29 $     &    $-816$     &  0.47   \\
$\eta_c$       &  2980   &   $2965$  &    679    &       75      &   $-123$     &    $-995$     &  0.31   \\
$J/\Psi$       &  3097   &   $3103$  &    488    &       97      &   $  29$     &    $-838$     &  0.35   \\
$B^0$          &  5280   &   $5261$  &    664    &       197     &   $ -34$     &    $-885$     &  0.50   \\
$B^*$          &  5325   &   $5305$  &    623    &       207     &   $  11$     &    $-855$     &  0.51   \\
$B_s^0$        &  5366   &   $5346$  &    612    &       143     &   $ -42$     &    $-868$     &  0.42   \\
$B_s^*$        &  5416   &   $5399$  &    555    &       155     &   $  13$     &    $-824$     &  0.44   \\
$B_c$          &  6277   &   $6244$  &    644    &       54      &   $-79$      &    $-1044$    &  0.26   \\
$B_c^*$        &  ...    &   $6336$  &    502    &       65      &   $  20$     &    $-921$     &  0.29   \\
$\eta_b$       &  9391   &   $9376$  &    740    &       24      &   $-96$      &    $-1305$    &  0.17   \\
$\Upsilon(1S)$ &  9460   &   $9486$  &    560    &       30      &   $  24$     &    $-1140$    &  0.19   \\
\toprule[0.8pt]
\end{tabular}
\end{table}

\subsection{Multiquark color flux-tube model (MCFTM)}

Details of the multiquark color flux-tube model based on traditional constituent quark models can be found in our previous paper~\cite{cscs}. Only prominent characteristics of the model are presented here. Within the framework of color flux-tube picture, the quark and antiquark in a meson are linked with a three-dimensional color flux tube, see Fig. 1. A two-body confinement potential can be written as
\begin{eqnarray}
V_{min}^{con}(2)=Kr_{ij}^2,
\end{eqnarray}
where $r_{ij}$ is distance between the quark and antiquark. The parameter $K$ is the stiffnesses of a three-dimension color flux-tube and determined by fitting the heavy-meson spectra. Comparing with the confinement potential in the CQM, one can obtain $K=-a_c\langle\mathbf{\lambda}^c_{i}\cdot\mathbf{\lambda}^c_{j}\rangle=800$ MeV fm$^{-2}$.

The fully-heavy tetraquark states favor the compact diquark-antiquark configuration rather than a loosely molecular states because of the lack of the light mesons exchange between two $Q\bar{Q}$-mesons. The color flux-tube structure of the diquark-antiquark configuration is given in Fig. 1.  According to the double Y-shaped color flux-tube structure, a four-body quadratic confinement potential can be written as,
\begin{eqnarray}
V^{con}(4)&=&K\left[ (\mathbf{r}_1-\mathbf{y}_{12})^2
+(\mathbf{r}_2-\mathbf{y}_{12})^2+(\mathbf{r}_{3}-\mathbf{y}_{34})^2\right. \nonumber \\
&+&
\left.(\mathbf{r}_4-\mathbf{y}_{34})^2+\kappa_d(\mathbf{y}_{12}-\mathbf{y}_{34})^2\right],
\end{eqnarray}
in which $\mathbf{r}_1$, $\mathbf{r}_2$, $\mathbf{r}_3$ and $\mathbf{r}_4$ represent the position of the $Q_1$, $Q_2$, $\bar{Q}_3$ and $\bar{Q}_4$, respectively. Two Y-shaped junctions $\mathbf{y}_{12}$ and $\mathbf{y}_{34}$ are variational parameters determined by taking the minimum of the confinement potential. The relative stiffness parameter $\kappa_{d}$ is equal to $\frac{C_{d}}{C_3}$~\cite{kappa}, where $C_{d}$ is the eigenvalue of the Casimir operator associated with the $SU(3)$ color representation $d$ at either end of the color flux-tube, such as $C_3=\frac{4}{3}$, $C_6=\frac{10}{3}$, and $C_8=3$.
\begin{figure}
\resizebox{0.48\textwidth}{!}{\includegraphics{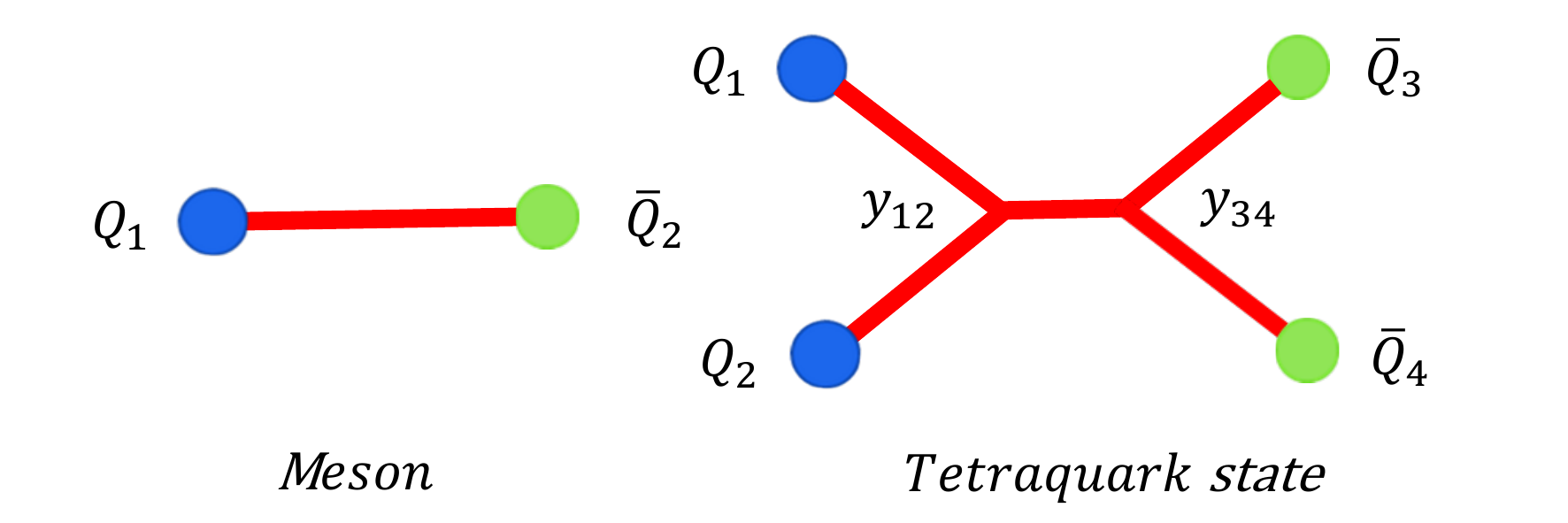}}
\caption{Color flux-tube structures.}
\label{flux-tube}
\end{figure}

The minimum of the confinement potential $V^{con}_{min}(4)$ can be obtained by taking the variation of $V^{con}(4)$ with respect to $\mathbf{y}_{12}$ and
$\mathbf{y}_{34}$, and it can be expressed as
\begin{eqnarray}
V^{con}_{min}(4)&=& K\left(\mathbf{R}_1^2+\mathbf{R}_2^2+
\frac{\kappa_{d}}{1+\kappa_{d}}\mathbf{R}_3^2\right),
\end{eqnarray}
The canonical coordinates $\mathbf{R}_i$ have the following forms,
\begin{eqnarray}
\mathbf{R}_{1} & = &
\frac{1}{\sqrt{2}}(\mathbf{r}_1-\mathbf{r}_2),~
\mathbf{R}_{2} =  \frac{1}{\sqrt{2}}(\mathbf{r}_3-\mathbf{r}_4), \nonumber \\
\mathbf{R}_{3} & = &\frac{1}{ \sqrt{4}}(\mathbf{r}_1+\mathbf{r}_2
-\mathbf{r}_3-\mathbf{r}_4), \\
\mathbf{R}_{4} & = &\frac{1}{ \sqrt{4}}(\mathbf{r}_1+\mathbf{r}_2
+\mathbf{r}_3+\mathbf{r}_4). \nonumber
\end{eqnarray}
The use of $V^{con}_{min}(4)$ can be understood here as that the gluon field readjusts immediately to its minimal configuration.

The OGE interaction is also involved in the MCFTM, which is the same as that of the CQM. It's worth mentioning that the MCFTM is not a completely new model but the updated version of the traditional CQM based on the color flux-tube picture of hadrons in the lattice QCD. In fact, it merely modifies the two-body confinement potential into the multibody one to describe multiquark states with multibody interaction. The MCFTM reduces to the CQM in the mesons while the MCFTM can obtain different results from the CQM in the multiquark states.

\section{wavefunction}

Numerical results of the states $[Q_1Q_2][\bar{Q}_3\bar{Q}_4]$ can be obtained by solving a fourbody Schr\"{o}dinger equation with their complete wavefunctions including all possible flavor-spin-color-spatial channels that contribute to a given well defined parity, isospin, and total angular momentum. In the frame of center-of-mass, the wavefunctions of the states $[Q_1Q_2][\bar{Q}_3\bar{Q}_4]$ can be constructed as a sum of the following direct products of color $\chi_c$, isospin $\eta_i$, spin $\chi_s$ and spatial $\phi^G_{lm}$ terms
\begin{eqnarray}
&&\Phi^{[Q_1Q_2][\bar{Q}_3\bar{Q}_4]}_{IM_IJM_J}=\sum_{\alpha}\xi_{\alpha}\left[\left[\left[\phi_{l_am_a}^G(\mathbf{r})\chi_{s_aM_{s_a}}\right]^{[Q_1Q_2]}_{J_aM_{J_a}}
\right.\right.\nonumber\\
&&\times\left.\left. \left[\phi_{l_bm_b}^G(\mathbf{R})
\chi_{s_bM_{s_b}}\right]^{[\bar{Q}_3\bar{Q}_4]}_{J_bM_{J_b}}\right ]_{J_{ab}M_{J_{ab}}}
\phi^G_{l_{ab}m_{ab}}(\mathbf{X})\right]_{JM_J}\nonumber\\
&&\times\left[\eta_{i_aM_{i_a}}^{[Q_1Q_2]}\eta_{i_bM_{i_b}}^{[\bar{Q}_3\bar{Q}_4]}\right]_{IM_I}
\left[\chi_{[c_a]W_{c_a}}^{[Q_1Q_2]}\chi_{[c_b]W_{c_b}}^{[\bar{Q}_3\bar{Q}_4]}\right]_{[C]W_C}
\end{eqnarray}
The subscripts $a$ and $b$ in the intermediate quantum numbers represent the $[Q_1Q_2]$ and $[\bar{Q}_3\bar{Q}_4]$, respectively. The summering index $\alpha$ stands for all possible flavor-spin-color-spatial intermediate quantum numbers. [~]'s denote Clebsh-Gordan coefficient coupling.

In the dynamical calculation, the relative spatial coordinates $\mathbf{r}$, $\mathbf{R}$ and $\mathbf{X}$ and center of mass $\mathbf{R}_c$ in the states $[Q_1Q_2][\bar{Q}_3\bar{Q}_4]$ can be defined as,
\begin{eqnarray}
\mathbf{r}&=&\mathbf{r}_1-\mathbf{r}_2,~~~\mathbf{R}=\mathbf{r}_3-\mathbf{r}_4 \nonumber\\
\mathbf{X}&=&\frac{m_1\mathbf{r}_1+m_2\mathbf{r}_2}{m_1+m_2}-\frac{m_3\mathbf{r}_3+m_4\mathbf{r}_4}{m_3+m_4},\\
\mathbf{R}_c&=&\frac{m_1\mathbf{r}_1+m_2\mathbf{r}_2+m_3\mathbf{r}_3+m_4\mathbf{r}_4}{m_1+m_2+m_3+m_4}.\nonumber
\end{eqnarray}
In the center-of-mass reference frame, the relative motion wave functions $\phi_{l_am_a}^G(\mathbf{r})$, $\phi_{l_bm_b}^G(\mathbf{R})$ and $\phi_{l_{ab}m_{ab}}^G(\mathbf{X})$ can
be expressed as the superposition of many different size Gaussian functions with well-defined quantum numbers, which share the exactly same form with that of the heavy mesons, to obtain accurate numerical results. For the sake of saving space, their explicit expressions are not presented here. The heavy quarks have isospin zero so they do not contribute to the total isospin. The flavor wavefunction is therefore symmetrical if $Q_1$ and $Q_2$ ($\bar{Q}_3$ and $\bar{Q}_4$) are identical particles.

The color representation of the $[\bar{Q}_3\bar{Q}_4]$ ($[Q_1Q_2]$) maybe antisymmetrical $\mathbf{3}_c$ ($\bar{\mathbf{3}}_c$) or symmetrical $\bar{\mathbf{6}}_c$ ($\mathbf{6}_c$). Coupling the colorful $[Q_1Q_2]$ and $[\bar{Q}_3\bar{Q}_4]$ into an overall colorless state according to color coupling rule, we have two different coupling ways: $\left[[Q_1Q_2]_{\bar{\mathbf{3}}_c}\otimes[\bar{Q}_3\bar{Q}_4]_{\mathbf{3}_c}\right]_{\mathbf{1}}$ and $\left[[Q_1Q_2]_{\mathbf{6}_c}\otimes[\bar{Q}_3\bar{Q}_4]_{\bar{\mathbf{6}}_c}\right]_{\mathbf{1}}$. The spin of the $[Q_1Q_2]$ is coupled to $s_a$ and that of the $[\bar{Q}_3\bar{Q}_4]$ is coupled to $s_b$. The total spin wavefunction of the state $[Q_1Q_2][\bar{Q}_3\bar{Q}_4]$ can be written as $S=s_a\oplus s_b$. Then we have the following basis vectors as a function of the total spin $S$, $0=1\oplus1~\mbox{or}~0\oplus0$, $1=1\oplus1,~1\oplus0~\mbox{or}~0\oplus1$, and $2=1\oplus1$.

Taking all degrees of freedom of identical particles in the $[Q_1Q_2]$ ($[\bar{Q}_3\bar{Q}_4]$) into account, the Pauli principle must be satisfied by imposing restrictions on their quantum numbers to satisfy antisymmetry. The S-wave $[Q_1Q_2]$ ($[\bar{Q}_3\bar{Q}_4]$) with two identical quarks (antiquarks) has two possible configurations,
$[Q_1Q_2]^1_{\bar{\mathbf{3}}_c}$ and $[Q_1Q_2]^0_{{\mathbf{6}}_c}$ ($[\bar{Q}_3\bar{Q}_4]^1_{\mathbf{3}_c}$ and $[\bar{Q}_3\bar{Q}_4]^0_{\bar{\mathbf{6}}_c}$),
where the superscript and subscript denote the spin and color representation, respectively. The possible color-flavor-spin functions of the states $[cc][\bar{c}\bar{c}]$, $[bb][\bar{c}\bar{c}]$ and $[bb][\bar{b}\bar{b}]$ states can be written as,
\begin{eqnarray}
0^+:&~&\left[[Q_1Q_2]^1_{\bar{\mathbf{3}}_c}[\bar{Q}_3\bar{Q}_4]^1_{\mathbf{3}_c}\right]^0_{\mathbf{1}_c},~  \left[[Q_1Q_2]^0_{{\mathbf{6}}_c}[\bar{Q}_3\bar{Q}_4]^0_{\bar{\mathbf{6}}_c}\right]^0_{\mathbf{1}_c} \nonumber\\
1^+:&~&\left[[Q_1Q_2]^1_{\bar{\mathbf{3}}_c}[\bar{Q}_3\bar{Q}_4]^1_{\mathbf{3}_c}\right]^1_{\mathbf{1}_c} \nonumber\\
2^+:&~&\left[[Q_1Q_2]^1_{\bar{\mathbf{3}}_c}[\bar{Q}_3\bar{Q}_4]^1_{\mathbf{3}_c}\right]^2_{\mathbf{1}_c}\nonumber
\end{eqnarray}
those of the states $[cc][\bar{c}\bar{b}]$ and $[bb][\bar{b}\bar{c}]$ reads,
\begin{eqnarray}
0^+:&~& \left[[Q_1Q_2]^1_{\bar{\mathbf{3}}_c}[\bar{Q}_3\bar{Q}_4]^1_{\mathbf{3}_c}\right]^0_{\mathbf{1}_c},
\left[[Q_1Q_2]^0_{{\mathbf{6}}_c}[\bar{Q}_3\bar{Q}_4]^0_{\bar{\mathbf{6}}_c}\right]^0_{\mathbf{1}_c}\nonumber\\
1^+:&~& \left[[Q_1Q_2]^1_{\bar{\mathbf{3}}_c}[\bar{Q}_3\bar{Q}_4]^{0,1}_{\mathbf{3}_c}\right]^1_{\mathbf{1}_c},
\left[[Q_1Q_2]^0_{{\mathbf{6}}_c}[\bar{Q}_3\bar{Q}_4]^1_{\bar{\mathbf{6}}_c}\right]^0_{\mathbf{1}_c}\nonumber\\
2^+:&~& \left[[Q_1Q_2]^1_{\bar{\mathbf{3}}_c}[\bar{Q}_3\bar{Q}_4]^1_{\mathbf{3}_c}\right]^2_{\mathbf{1}_c}\nonumber
\end{eqnarray}
those of the state $[cb][\bar{c}\bar{b}]$ reads,
\begin{eqnarray}
0^+:&~& \left[[Q_1Q_2]^{0,1}_{\bar{\mathbf{3}}_c}[\bar{Q}_3\bar{Q}_4]^{0,1}_{\mathbf{3}_c}\right]^0_{\mathbf{1}_c},
\left[[Q_1Q_2]^{0,1}_{{\mathbf{6}}_c}[\bar{Q}_3\bar{Q}_4]^{0,1}_{\bar{\mathbf{6}}_c}\right]^0_{\mathbf{1}_c}\nonumber\\
1^+:&~& \left[[Q_1Q_2]^{0,1}_{\bar{\mathbf{3}}_c}[\bar{Q}_3\bar{Q}_4]^{0,1}_{\mathbf{3}_c}\right]^1_{\mathbf{1}_c},
\left[[Q_1Q_2]^0_{{\mathbf{6}}_c}[\bar{Q}_3\bar{Q}_4]^{0,1}_{\bar{\mathbf{6}}_c}\right]^{1}_{\mathbf{1}_c}\nonumber\\
2^+:&~& \left[[Q_1Q_2]^1_{\bar{\mathbf{3}}_c}[\bar{Q}_3\bar{Q}_4]^1_{\mathbf{3}_c}\right]^2_{\mathbf{1}_c}, \left[[Q_1Q_2]^1_{\mathbf{6}_c}[\bar{Q}_3\bar{Q}_4]^1_{\bar{\mathbf{6}}_c}\right]^2_{\mathbf{1}_c}\nonumber
\end{eqnarray}

In the following, we will extend the three models to study the properties of the states $[Q_1Q_2][\bar{Q}_3\bar{Q}_4]$ with the well-defined wavefunction.

\section{numerical results and discussions}

\subsection{Fully-heavy tetraquark states}

Recently, various versions of color-magnetic interaction models were widely utilized to investigate the properties of the states $[Q_1Q_2][\bar{Q}_3\bar{Q}_4]$. Berezhnoy et al applied a color-magnetic model, in which the tetraquark mass can be determined by solving a two-particle Schrodinger equation with the pointlike diquark (antidiquark) in color $\bar{\mathbf{3}}_c$ ($\mathbf{3}_c$), to research the states $[cc][\bar{c}\bar{c}]$, $[bb][\bar{b}\bar{b}]$ and $[bc][\bar{b}\bar{c}]$. With the exception of the tensor states $[cc][\bar{c}\bar{c}]$ and $[bc][\bar{b}\bar{c}]$, the lowest states with other quantum numbers are all below relevant two meson thresholds~\cite{berezhnoy}. Karliner et al. studied the $0^{++}$ states $[cc][\bar{c}\bar{c}]$ and $[bb][\bar{b}\bar{b}]$ with the color-magnetic interaction model motivated by the QCD-string junction picture~\cite{karliner}. Their masses are, respectively, $6192\pm25$ MeV and $18826\pm25$ MeV. It was noted that an experimental search for these states in the predicted mass range is highly desirable.
Wu et al systematically investigated the mass spectra of the states $[Q_1Q_2][\bar{Q}_3\bar{Q}_4]$ with a color-magnetic interaction model with a reference mass scale~\cite{cmwj}.
It was found that the states $[bb][\bar{b}\bar{c}]$ and $[bc][\bar{b}\bar{c}]$ are possible stable or narrow resonance states.

\begin{table*}
\caption{The mass spectra of the ground states $[Q_1Q_2][\bar{Q}_3\bar{Q}_4]$ in the three models, unit in MeV.}\label{4Q}
\begin{tabular}{ccccccccccccccc}
\toprule[0.8pt]\noalign{\smallskip}
Model&&&CMIM&&&MCFTM&&&CQM\\
\toprule[0.8pt]\noalign{\smallskip}
Flavor               &~$J^{P}$~&~~~$\bar{\mathbf{3}}_c\otimes\mathbf{3}_c$~~~&~~~$\mathbf{6}_c\otimes\bar{\mathbf{6}}_c$~~~&~~~~C.C.~~~~&  $~~~~\bar{\mathbf{3}}_c\otimes\mathbf{3}_c$~~~~&~~~~$\mathbf{6}_c\otimes\bar{\mathbf{6}}_c$~~~~&~~C.C.~~&  $~~~~\bar{\mathbf{3}}_c\otimes\mathbf{3}_c$~~~~&~~~~$\mathbf{6}_c\otimes\bar{\mathbf{6}}_c$~~~~&~C.C.~ \\
\toprule[0.8pt]\noalign{\smallskip}
                       &  $0^{+}$   & $-28.27$,~$66\%$  &  42.40,~$34\%$   &  $-102.64$,~6035 &  6454,~$56\%$   & 6467,~$44\%$  & 6407  & 6573,~$36\%$   &  6537,~$64\%$ & 6491  \\
$[cc][\bar{c}\bar{c}]$ &  $1^{+}$   & ~~0.00,~$100\%$   &     ...          &  0.00,~6139      &  6463,~$100\%$  &   ...         & 6463  & 6580,~$100\%$  &     ...       & 6580  \\
                       &  $2^{+}$   & 56.53,~$100\%$    &     ...          &  56.53,~6194     &  6486,~$100\%$  &   ...         & 6486  & 6607,~$100\%$  &     ...       & 6607  \\
\noalign{\smallskip}
                       &  $0^{+}$   & $-13.33$,~$66\%$  &  32.80,~$34\%$   &  $-58.92$,~12597 &  12940,~$49\%$  & 12938,~$51\%$ & 12906 & 13023,~$29\%$  & 12986,~$71\%$ & 12963 \\
$[cc][\bar{b}\bar{b}]$ &  $1^{+}$   & 4.27,~$100\%$     &      ...         &  4.27,~12660     &  12945,~$100\%$ &    ...        & 12945 & 13024,~$100\%$ &     ...       & 13024 \\
                       &  $2^{+}$   & 39.47,~$100\%$    &     ...          &  39.47,~12695    &  12960,~$100\%$ &    ...        & 12960 & 13041,~$100\%$ &     ...       & 13041 \\
\noalign{\smallskip}
                       &  $0^{+}$   & $-15.47$,~$66\%$  &  23.20,~$34\%$   &  $-56.16$,~18834 &  19377,~$38\%$  & 19351,~$62\%$ & 19329 & 19417,~$28\%$  & 19368,~$72\%$ & 19357 \\
$[bb][\bar{b}\bar{b}]$ &  $1^{+}$   & 0.00,~$100\%$     &      ...         &  0.00,~18890     &  19373,~$100\%$ &    ...        & 19373 & 19413,~$100\%$ &      ...      & 19413 \\
                       &  $2^{+}$   & 30.93,~$100\%$    &      ...         &  30.93,~18921    &  19387,~$100\%$ &    ...        & 19387 & 19429,~$100\%$ &      ...      & 19429 \\
\noalign{\smallskip}
                       &  $0^+$     & $-22.93$,~$66\%$  &  34.40,~$34\%$   &  $-83.27$,~9314  &  9705,~$56\%$   & 9721,~$44\%$  & 9670  & 9813,~$41\%$   &  9780,~$59\%$ & 9753  \\
$[cc][\bar{c}\bar{b}]$ &  $1^+$     & $-15.85$,~$65\%$  &  16.80,~$35\%$   &  $-53.17$,~9343  &  9705,~$58\%$   & 9712,~$42\%$  & 9683  & 9808,~$31\%$   &  9785,~$69\%$ & 9766  \\
                       &  $2^+$     & 45.87,~$100\%$    &      ...         &  45.87,~9442     &  9732,~$100\%$  &    ...        & 9732  & 9839,~$100\%$  &     ...       & 9839  \\
\noalign{\smallskip}
                       &  $0^+$     & $-16.53$,~$66\%$  &  24.80,~$34\%$   &  $-60.03$,~15713 &  16158,~$42\%$  & 16158,~$58\%$ & 16126 & 16224,~$31\%$  & 16201,~$69\%$ & 16175 \\
$[bb][\bar{c}\bar{b}]$ &  $1^+$     & $-18.79$,~$67\%$  &  7.20,~$33\%$    &  $-43.33$,~15729 &  16151,~$39\%$  & 16139,~$61\%$ & 16130 & 16230,~$23\%$  & 16187,~$77\%$ & 16179 \\
                       &  $2^+$     & 33.07,~$100\%$    &      ...         &   33.07,~15806   &  16182,~$100\%$ &    ...        & 16182 & 16274,~$100\%$ &      ...      & 16274 \\
\noalign{\smallskip}
                       &  $0^{+}$   & $-53.24$,~$33\%$  &$-108.10$,~$67\%$ &  $-159.37$,~12354&  12955,~$33\%$  & 12898,~$67\%$ & 12829 & 13043,~$29\%$  & 12968,~$71\%$ & 12894 \\
$[cb][\bar{c}\bar{b}]$ &  $1^{+}$   & $-21.81$,~$28\%$  & $-62.04$,~$72\%$ &  $-77.75$,~12436 &  12955,~$40\%$  & 12938,~$60\%$ & 12881 & 13052,~$33\%$  & 13006,~$67\%$ & 12955 \\
                       &  $2^{+}$   & 34.13,~$33\%$     &  43.73,~$67\%$   &  34.13,~12548    &  12984,~$36\%$  & 12959,~$64\%$ & 12925 & 13084,~$27\%$  & 13032,~$73\%$ & 13000 \\
\toprule[0.8pt]\noalign{\smallskip}

\end{tabular}
%\end{table*}
%\begin{table*}[ht]
\caption{The values of various parts of the Hamiltonian in the MCFTM, unit in MeV.}\label{part}
\begin{tabular}{ccccccccccccccccccc}
\toprule[0.8pt] \noalign{\smallskip}
Flavor  &~~$J^{P}$~~&~~~$E_4$~~~ &~~~$\langle E_k\rangle$~~~& $\langle V^{con}_{min}(4)\rangle$ &~~~$\langle V^{cm}\rangle$~~~ & $\langle V^{clb}\rangle$ & ~~~~~~$T_{M_1M_2}$~~~~~~ &~$\Delta E$~ &~~~$\Delta\langle E_k\rangle$~~~&
$\Delta \langle V^{con}_{min}(4)\rangle$ & $\Delta\langle V^{cm}\rangle$ & $\Delta\langle V^{clb}\rangle$    \\
\noalign{\smallskip}
\toprule[0.8pt] \noalign{\smallskip}
                       &  $0^{+}$  & 6407  & 887 & 192 & $-51$ & $-1279$ & $\eta_c\eta_c$             & 477 & $-471$ & 42 & 195    & 711   \\
$[cc][\bar{c}\bar{c}]$ &  $1^{+}$  & 6463  & 800 & 203 &   4   & $-1202$ & $\eta_c\Psi$               & 395 & $-367$ & 32 & 98     & 632   \\
                       &  $2^{+}$  & 6486  & 769 & 211 &   27  & $-1178$ & $\Psi\Psi$                 & 280 & $-206$ & 18 & $-31$  & 499   \\
\noalign{\smallskip}
                       &  $0^{+}$  & 12906 & 853 & 131 & $-27$ & $-1392$ & $B_cB_c$                   & 418 & $-435$ & 24 & 132    & 696   \\
$[cc][\bar{b}\bar{b}]$ &  $1^{+}$  & 12945 & 787 & 135 &  6    & $-1324$ & $B_c^*B_c$                 & 365 & $-359$ & 17 & 65     & 642   \\
                       &  $2^{+}$  & 12960 & 764 & 139 &  20   & $-1304$ & $B_c^*B_c^*$               & 288 & $-240$ & 9  & $-20$  & 538   \\
\noalign{\smallskip}
                       &  $0^{+}$  & 19329 & 865 & 69  & $-26$ & $-1605$ & $\eta_b\eta_b$             & 577 & $-615$ & 21 & 166    & 1005  \\
$[bb][\bar{b}\bar{b}]$ &  $1^{+}$  & 19373 & 826 & 68  &  3    & $-1550$ & $\eta_b\Upsilon(1S)$       & 511 & $-474$ & 14 & 75     & 895   \\
                       &  $2^{+}$  & 19387 & 799 & 70  &  17   & $-1525$ & $\Upsilon(1S)\Upsilon(1S)$ & 415 & $-321$ & 10 & $-31$  & 756   \\
\noalign{\smallskip}
                       &  $0^{+}$  & 9670  & 858 & 161 & $-38$ & $-1309$ & $\eta_cB_c$                & 461 & $-465$ & 32 & 164    & 730   \\
$[cc][\bar{c}\bar{b}]$ &  $1^{+}$  & 9683  & 838 & 165 & $-25$ & $-1295$ & $\eta_cB_c^*$              & 382 & $-343$ & 25 & 78     & 621   \\
                       &  $2^{+}$  & 9732  & 758 & 174 &   22  & $-1221$ & $\Psi B_c^*$               & 293 & $-232$ & 12 & $-27$  & 538   \\
\noalign{\smallskip}
                       &  $0^{+}$  & 16126 & 856 & 97  & $-27$ & $-1483$ & $B_c\eta_b$                & 506 & $-528$ & 19 & 148    & 866   \\
$[bb][\bar{c}\bar{b}]$ &  $1^{+}$  & 16130 & 905 & 94  & $-19$ & $-1530$ & $B_c^*\eta_b$              & 418 & $-337$ & 6  & 58     & 696   \\
                       &  $2^{+}$  & 16182 & 771 & 102 &  17   & $-1392$ & $B_c^*\Upsilon(1S)$        & 360 & $-291$ & 8  & $-27$  & 669   \\
\noalign{\smallskip}
                       &  $0^{+}$  & 12829 & 932 & 123 & $-84$ & $-1483$ & $\eta_b\eta_c$             & 344 & $-487$ & 24 & 135    & 817   \\
$[cb][\bar{c}\bar{b}]$ &  $1^{+}$  & 12881 & 816 & 134 & $-31$ & $-1379$ & $\eta_c\Upsilon(1S)$       & 430 & $-423$ & 29 & 68     & 756   \\
                       &  $2^{+}$  & 12925 & 789 & 144 &  21   & $-1370$ & $\Psi\Upsilon(1S)$         & 336 & $-259$ & 18 & $-32$  & 608   \\
\toprule[0.8pt]
\end{tabular}
%\end{table*}
%\begin{table*}[ht]
\caption{The average distances $\langle\mathbf{r}_{ij}^2\rangle^{\frac{1}{2}}$ and $\langle\mathbf{X}^2\rangle^{\frac{1}{2}}$ of the ground states $[Q_1Q_2][\bar{Q}_3\bar{Q}_4]$ in the MCFTM, unit in fm.}\label{rms}
\begin{tabular}{cccccccccccccccccccccccccccc}
\toprule[0.8pt] \noalign{\smallskip}
State &&&$[cc][\bar{c}\bar{c}]$ &&&& $[cc][\bar{b}\bar{b}]$ &&&& $[bb][\bar{b}\bar{b}]$ &&&& $[cc][\bar{c}\bar{b}]$ &&&& $[bb][\bar{c}\bar{b}]$ &&&& $[bc][\bar{b}\bar{c}]$   \\
\noalign{\smallskip}
$J^{P}$ &&$0^{+}$&$1^{+}$&$2^{+}$&~~~&$0^{+}$&$1^{+}$&$2^{+}$&~~~&$0^{+}$&$1^{+}$&$2^{+}$&~~~&$0^{+}$&$1^{+}$&$2^{+}$&~~~&$0^{+}$&$1^{+}$&$2^{+}$&~~~&$0^{+}$&$1^{+}$&$2^{+}$ \\
$\langle\mathbf{r}_{12}^2\rangle^{\frac{1}{2}}$ &&0.44&0.44&0.44& &0.42&0.42&0.42& &0.27&0.25&0.26& &0.43&0.43&0.43& &0.28&0.28&0.27& &0.37&0.38&0.40 \\
\noalign{\smallskip}
$\langle\mathbf{r}_{34}^2\rangle^{\frac{1}{2}}$ &&0.44&0.44&0.44& &0.30&0.28&0.28& &0.27&0.25&0.26& &0.38&0.38&0.38& &0.35&0.35&0.35& &0.37&0.38&0.40 \\
\noalign{\smallskip}
$\langle\mathbf{r}_{13}^2\rangle^{\frac{1}{2}}$ &&0.44&0.47&0.48& &0.36&0.39&0.39& &0.26&0.27&0.28& &0.45&0.46&0.48& &0.36&0.36&0.38& &0.24&0.26&0.26 \\
\noalign{\smallskip}
$\langle\mathbf{r}_{24}^2\rangle^{\frac{1}{2}}$ &&0.44&0.47&0.48& &0.36&0.39&0.39& &0.26&0.27&0.28& &0.36&0.37&0.40& &0.26&0.26&0.29& &0.44&0.46&0.47 \\
\noalign{\smallskip}
$\langle\mathbf{r}_{14}^2\rangle^{\frac{1}{2}}$ &&0.44&0.47&0.48& &0.36&0.39&0.39& &0.26&0.27&0.28& &0.36&0.37&0.40& &0.26&0.26&0.29& &0.35&0.37&0.38 \\
\noalign{\smallskip}
$\langle\mathbf{r}_{23}^2\rangle^{\frac{1}{2}}$ &&0.44&0.47&0.48& &0.36&0.39&0.39& &0.26&0.27&0.28& &0.45&0.46&0.48& &0.36&0.36&0.38& &0.35&0.37&0.38 \\
\noalign{\smallskip}
$\langle\mathbf{X}^2\rangle^{\frac{1}{2}}$      &&0.31&0.36&0.37& &0.25&0.29&0.30& &0.17&0.20&0.21& &0.28&0.29&0.33& &0.20&0.20&0.24& &0.20&0.22&0.22 \\
\toprule[0.8pt]
\end{tabular}
\end{table*}

\begin{table*}[ht]
\caption{The average values of various parts of the Hamiltonian in MeV, the average distances in fm of the states $[cc][\bar{c}\bar{c}]$, $[cc][\bar{b}\bar{b}]$ and $[bb][\bar{b}\bar{b}]$ in the MCFTM.} \label{proportion}
\begin{tabular}{cccccccccccccccccc}
\toprule[0.8pt] \noalign{\smallskip}
&$LS$&~$J^{P}$~&~States~&~Mass,~prop.~&~$\langle E_k\rangle$~& $\langle V^{com}_{min}(4)\rangle$ &  $\langle V^{cm}\rangle$  & $\langle V^{clb}\rangle$  &~~$\langle\mathbf{r}_{12}^2\rangle^{\frac{1}{2}}$~~&$\langle\mathbf{r}_{34}^2\rangle^{\frac{1}{2}}$&~$\langle\mathbf{r}_{13}^2\rangle^{\frac{1}{2}}$~&
$\langle\mathbf{r}_{24}^2\rangle^{\frac{1}{2}}$&~$\langle\mathbf{r}_{14}^2\rangle^{\frac{1}{2}}$~&$\langle\mathbf{r}_{23}^2\rangle^{\frac{1}{2}}$&
~$\langle\mathbf{X}^2\rangle^{\frac{1}{2}}$ \\
\noalign{\smallskip}
\toprule[0.8pt]
\noalign{\smallskip}
                 &  &             & $\bar{\mathbf{3}}_c\otimes\mathbf{3}_c$ & 6454,~56\%     & 878  & 188 & $-11$ & $-1258$ & 0.42 & 0.42 & 0.45 & 0.45 & 0.45 & 0.45 & 0.33  \\
                 &00&$0^{+}$      & $\mathbf{6}_c\otimes\bar{\mathbf{6}}_c$ & 6467,~44\%     & 899  & 199 & 17    & $-1306$ & 0.46 & 0.46 & 0.43 & 0.43 & 0.43 & 0.43 & 0.28  \\
                 &  &             & C.C.                                    & 6407           & 887  & 192 & $-51$ & $-1279$ & 0.44 & 0.44 & 0.44 & 0.44 & 0.44 & 0.44 & 0.31  \\
\noalign{\smallskip}
                 &  &             & $\bar{\mathbf{3}}_c\otimes\mathbf{3}_c$ & 6730,~98\%     & 783  & 283 & 4     & $-997$  & 0.47 & 0.47 & 0.61 & 0.61 & 0.61 & 0.61 & 0.52  \\
$[cc][\bar{c}\bar{c}]$&10&$1^{-}$ & $\mathbf{6}_c\otimes\bar{\mathbf{6}}_c$ & 6888,~2\%      & 910  & 274 & 12    & $-966$  & 0.51 & 0.51 & 0.54 & 0.54 & 0.54 & 0.54 & 0.40  \\
                 &  &             & C.C.                                    & 6727           & 785  & 283 & $-2$  & $-997$  & 0.47 & 0.47 & 0.61 & 0.61 & 0.61 & 0.61 & 0.51  \\
\noalign{\smallskip}
                 &  &             & $\bar{\mathbf{3}}_c\otimes\mathbf{3}_c$ & 6995,~$>$99\%  & 802  & 364 & 9     & $-888$  & 0.48 & 0.48 & 0.75 & 0.75 & 0.75 & 0.75 & 0.66  \\
                 &20&$2^{+}$      & $\mathbf{6}_c\otimes\bar{\mathbf{6}}_c$ & 7213,~$<$1\%   & 978  & 339 & 10    & $-772$  & 0.55 & 0.55 & 0.63 & 0.63 & 0.63 & 0.63 & 0.50  \\
                 &  &             & C.C.                                    & 6944           & 802  & 364 & 8     & $-887$  & 0.48 & 0.48 & 0.75 & 0.75 & 0.75 & 0.75 & 0.66  \\
\noalign{\smallskip}\noalign{\smallskip}
                 &  &             & $\bar{\mathbf{3}}_c\otimes\mathbf{3}_c$ & 12939,~41\%    & 847  & 127 & $-3$  & $-1372$ & 0.27 & 0.41 & 0.37 & 0.37 & 0.37 & 0.37 & 0.29  \\
                 &00&$0^{+}$      & $\mathbf{6}_c\otimes\bar{\mathbf{6}}_c$ & 12938,~51\%    & 859  & 135 & 13    & $-1411$ & 0.33 & 0.42 & 0.35 & 0.35 & 0.35 & 0.35 & 0.23  \\
                 &  &             & C.C.                                    & 12906          & 853  & 131 & $-27$ & $-1392$ & 0.30 & 0.42 & 0.36 & 0.36 & 0.36 & 0.36 & 0.25  \\
\noalign{\smallskip}
                 &  &             & $\bar{\mathbf{3}}_c\otimes\mathbf{3}_c$ & 13204,~$>$99\% & 727  & 201 & 6     & $-1071$ & 0.30 & 0.46 & 0.52 & 0.52 & 0.52 & 0.52 & 0.45  \\
$[bb][\bar{c}\bar{c}]$&10&$1^{-}$ & $\mathbf{6}_c\otimes\bar{\mathbf{6}}_c$ & 13370,~$<$1\%  & 884  & 186 & 9     & $-1051$ & 0.36 & 0.48 & 0.44 & 0.44 & 0.44 & 0.44 & 0.32  \\
                 &  &             & C.C.                                    & 13204          & 728  & 201 & 4     & $-1071$ & 0.30 & 0.46 & 0.52 & 0.52 & 0.52 & 0.52 & 0.45  \\
\noalign{\smallskip}
                 &  &             & $\bar{\mathbf{3}}_c\otimes\mathbf{3}_c$ & 13398,~$>$99\% & 727  & 267 & 8     & $-946$  & 0.31 & 0.48 & 0.65 & 0.65 & 0.65 & 0.65 & 0.58  \\
                 &20&$2^{+}$      & $\mathbf{6}_c\otimes\bar{\mathbf{6}}_c$ & 13696,~$<$1\%  & 954  & 235 & 7     & $-842$  & 0.36 & 0.53 & 0.52 & 0.52 & 0.52 & 0.52 & 0.40  \\
                 &  &             & C.C.                                    & 13398          & 727  & 267 & 8     & $-946$  & 0.31 & 0.48 & 0.65 & 0.65 & 0.65 & 0.65 & 0.58  \\
\noalign{\smallskip}\noalign{\smallskip}
                 &  &             & $\bar{\mathbf{3}}_c\otimes\mathbf{3}_c$ & 19367,~38\%    & 899  & 63  & $-6$  & $-1615$ & 0.24 & 0.24 & 0.26 & 0.26 & 0.26 & 0.26 & 0.19  \\
                 &00&$0^{+}$      & $\mathbf{6}_c\otimes\bar{\mathbf{6}}_c$ & 19352,~62\%    & 884  & 72  & 9     & $-1638$ & 0.28 & 0.28 & 0.25 & 0.25 & 0.25 & 0.25 & 0.16  \\
                 &  &             & C.C.                                    & 19329          & 865  & 69  & $-26$ & $-1605$ & 0.27 & 0.27 & 0.26 & 0.26 & 0.26 & 0.26 & 0.17  \\
\noalign{\smallskip}
                 &  &             & $\bar{\mathbf{3}}_c\otimes\mathbf{3}_c$ & 19636,~$>$99\% & 700  & 110 & 4     & $-1204$ & 0.29 & 0.29 & 0.39 & 0.39 & 0.39 & 0.39 & 0.33  \\
$[bb][\bar{b}\bar{b}]$&10&$1^{-}$ & $\mathbf{6}_c\otimes\bar{\mathbf{6}}_c$ & 19792,~$<$1\%  & 854  & 104 & 6     & $-1198$ & 0.32 & 0.32 & 0.32 & 0.32 & 0.32 & 0.32 & 0.23  \\
                 &  &             & C.C.                                    & 19635          & 701  & 110 & 2     & $-1204$ & 0.29 & 0.29 & 0.39 & 0.39 & 0.39 & 0.39 & 0.33  \\
\noalign{\smallskip}
                 &  &             & $\bar{\mathbf{3}}_c\otimes\mathbf{3}_c$ & 19812,~$>$99\% & 659  & 157 & 6     & $-1035$ & 0.31 & 0.31 & 0.50 & 0.50 & 0.50 & 0.50 & 0.45  \\
                 &20&$2^{+}$      & $\mathbf{6}_c\otimes\bar{\mathbf{6}}_c$ & 20105,~$<$1\%  & 898  & 136 & 4     & $-960$  & 0.36 & 0.36 & 0.39 & 0.39 & 0.39 & 0.39 & 0.29  \\
                 &  &             & C.C.                                    & 19812          & 659  & 157 & 6     & $-1035$ & 0.31 & 0.31 & 0.50 & 0.50 & 0.50 & 0.50 & 0.45  \\
\noalign{\smallskip}
\toprule[0.8pt]
\end{tabular}
\end{table*}

One should note that all color-magnetic interaction models ignore the spatial degree of freedom so that everything in the models depends only on the color-spin algebra. The generalization of the color-magnetic interaction models from the conventional hadrons to the multiquark states is implemented under the assumption that the spatial configurations of
each $qq$, $q\bar{q}$ and $\bar{q}\bar{q}$ pairs are the same in the multiquark states as in the ordinary hadrons. The well-known H-particle predicted in the color-magnetic interaction model was below the $\Lambda\Lambda$ threshold about 80 MeV~\cite{h-particle}. However, the state was above the threshold in the nonrelativistic quark model involving the color-magnetic interaction with spatial degree of freedom and other various dynamics once $SU(3)$ flavor symmetry is broken~\cite{h-nrqm}. The state was once very fashionable and searched for in many experiments. The high-sensitivity search at Brookhaven gave no evidence for the production of the state~\cite{exp-hparticle}. Recently, the theoretical case for the state continues to be strong and has been strengthened by the NPLQCD and HALQCD collaborations that both observed the state~\cite{nplqcd,halqcd}. The high-statistics search for the state production shown that no indication of the state with a mass near the $\Lambda\Lambda$ threshold was seen~\cite{exp-hparticle2}.

In view of the inherent defects of the color-magnetic models and the experience of the H-particle, it is therefore necessary to make an systematically dynamical investigation on the properties of the states $[Q_1Q_2][\bar{Q}_3\bar{Q}_4]$ in the quark models containing various QCD dynamical effects. The MCFTM and CQM are therefore involved, in which the masses of all possible states $[Q_1Q_2][\bar{Q}_3\bar{Q}_4]$ can be obtained by solving a four-body Schr\"{o}dinger equation with the well-defined trial wavefunctions and presented in Table~\ref{4Q}. The notations $\bar{\mathbf{3}}_c\otimes\mathbf{3}_c$ and $\mathbf{6}_c\otimes\bar{\mathbf{6}}_c$ stand for the color configurations $\left[[Q_1Q_2]_{\bar{\mathbf{3}}_c}[\bar{Q}_2\bar{Q}_4]_{\mathbf{3}_c}\right]_{\mathbf{1}}$ and  $\left[[Q_1Q_2]_{\mathbf{6}_c}[\bar{Q}_3\bar{Q}_4]_{\bar{\mathbf{6}}_c}\right]_{\mathbf{1}}$, respectively. C.C. represents the coupling of the two color configurations. The
masses of the states $[Q_1Q_2][\bar{Q}_3\bar{Q}_4]$ with two color configurations and their individual proportion in the eigen states can be achieved by the eigen wavefunction,
which are listed in Table~\ref{4Q}. In order to facilitate the comparison, we also reproduce the color-magnetic interaction energy of the states in the CMIM with the approximation $C_{QQ}=C_{Q\bar{Q}}$~\cite{cmwj}. The proportion of each color configuration and the mass (right) obtained with the reference mass formula in the coupling results are given.

It can be found from Table~\ref{4Q} that the masses predicted by the CMIM are lower 300-500 MeV than those predicted by other two models involving QCD dynamic effects. The masses predicted by the MCFTM are lower 30-120 MeV than those by the CQM. Comparing the masses with the lowest two meson thresholds $T_{M_1M_2}$, the binding energy $\Delta E$ in the MCFTM can be calculated and are presented in Table~\ref{part}. One can find that none of states can exist as a bound state because all states are hundreds of MeV above the corresponding threshold in the MCFTM while the masses of the states predicted by the CMIM are close to the corresponding threshold. In order to unveil the underlying cause, the average values of various parts in the Hamiltonian are given by using the eigen wavefunction in Table~\ref{part}, in which $\langle E_k\rangle$, $\langle V^{con}\rangle$, $\langle V^{cm}\rangle$ and $\langle V^{clb}\rangle$ represent the average values of kinetic energy, confinement potential, color-magnetic interaction and Coulomb interaction, respectively. The $\Delta$ stands for the corresponding interaction difference between the state $[Q_1Q_2][\bar{Q}_3\bar{Q}_4]$ and two-meson thresholds. The average distances $\langle\mathbf{r}_{ij}^2\rangle^{\frac{1}{2}}$ between the particles $i$ and $j$ and $\langle\mathbf{X}^2\rangle^{\frac{1}{2}}$ between the $[Q_1Q_2]$ and $[\bar{Q}_3\bar{Q}_4]$ are presented in Table~\ref{rms}.

The investigation on the spectrum of the heavy-mesons in Sec. II indicates that the Coulomb interaction is significant in the formation of the heavy-mesons. It can be found from Table~\ref{part} that the interaction also plays a decisive role in the states $[Q_1Q_2][\bar{Q}_3\bar{Q}_4]$. The Coulomb interaction provides very strong attraction in the heavy-mesons and the states $[Q_1Q_2][\bar{Q}_3\bar{Q}_4]$. The interaction depends on $\frac{1}{r}$ and the color factor $\langle\mathbf{\lambda}^c_i\cdot\mathbf{\lambda}^c_j\rangle$, the strength of which is related to the color factor $\langle\mathbf{\lambda}^c_i\cdot\mathbf{\lambda}^c_j\rangle$.
In the heavy quark sector, the large quark mass allows two particles to approach each other as a result of a small kinetic, which helps to strengthen the Coulomb interaction. In
the heavy mesons, $\langle\mathbf{\lambda}^c_i\cdot\mathbf{\lambda}^c_j\rangle=-\frac{16}{3}$, which is stronger than those of the states $[Q_1Q_2][\bar{Q}_3\bar{Q}_4]$, see Table~\ref{color-matrix}. The states $[Q_1Q_2][\bar{Q}_3\bar{Q}_4]$ are therefore looser than heavy mesons $\eta_c$, $\Psi$, $B_c$, $\eta_b$ and $\Upsilon$, see the average distances in Table~\ref{spectra} and~\ref{rms}. In addition, the values of the Coulomb interaction in the states $[Q_1Q_2][\bar{Q}_3\bar{Q}_4]$ are higher 500-1000 MeV than those of their corresponding two-meson thresholds, see the difference $\Delta\langle V^{clb}\rangle$ in Table~\ref{part}, which is main reason resulting in none of bound states in the quark models with QCD dynamic effects. It is therefore difficult for the CMIM to completely absorb the strong Coulomb interaction effect in the states $[Q_1Q_2][\bar{Q}_3\bar{Q}_4]$ by the effective constituent quark mass.
\begin{table}[ht]
\caption{Color matrix elements, $\hat{O}_{ij}=\mathbf{\lambda}^c_i\cdot\mathbf{\lambda}^c_j$.} \label{color-matrix}
\begin{tabular}{ccccccccccc}
\toprule[0.8pt] \noalign{\smallskip}
$\langle\hat{O}_{ij}\rangle$&$\langle\hat{O}_{12}\rangle$&$\langle\hat{O}_{34}\rangle$&$\langle\hat{O}_{13}\rangle$&$\langle\hat{O}_{24}\rangle$&$\langle\hat{O}_{14}\rangle$& $\langle\hat{O}_{23}\rangle$   \\
\noalign{\smallskip}
\toprule[0.8pt]
\noalign{\smallskip}
$\langle\bar{\mathbf{3}}_c\otimes\mathbf{3}_c|\hat{O}_{ij}|\bar{\mathbf{3}}_c\otimes\mathbf{3}_c\rangle$&$-\frac{8}{3}$&$-\frac{8}{3}$&$-\frac{4}{3}$&$-\frac{4}{3}$&$-\frac{4}{3}$&$-\frac{4}{3}$\\
 \noalign{\smallskip}
$\langle\mathbf{6}_c\otimes\bar{\mathbf{6}}_c|\hat{O}_{ij}|\mathbf{6}_c\otimes\bar{\mathbf{6}}_c\rangle$&$\frac{4}{3}$&$\frac{4}{3}$&$-\frac{10}{3}$&$-\frac{10}{3}$&$-\frac{10}{3}$&$-\frac{10}{3}$\\
 \noalign{\smallskip}
$\langle\bar{\mathbf{3}}_c\otimes\mathbf{3}_c|\hat{O}_{ij}|\mathbf{6}_c\otimes\bar{\mathbf{6}}_c\rangle$&$0$&$0$&$-2\sqrt{2}$&$-2\sqrt{2}$&$2\sqrt{2}$&$2\sqrt{2}$\\
 \noalign{\smallskip}
\toprule[0.8pt]
\end{tabular}
\end{table}

The long-range confinement interaction contributes a little to the masses and binding energy of the ground states $[Q_1Q_2][\bar{Q}_3\bar{Q}_4]$ because of the small distances, see the average values $V^{con}_{min}$ and $\Delta V^{con}_{min}$ in Table~\ref{part}. The mass difference, about 30-120 MeV, bwtween the CQM and MCFTM in Table~\ref{4Q} originates from different types of confinement potential. The multibody confinement potential based on the lattice color flux-tube picture is thought to be closer to real physical images than two-body one related to color charges, which plays significant roles in many interesting places of hadron physics, such as the formation and decay of the multiquark states via strong interaction, quark pair creation and hadron structure. The multibody confinement potential can reduce the mass of the multiquark states. Similar quark models with different type of multibody confinement potential have been extensively applied to study the properties of the multiquark states~\cite{multibody}.

The CMIM does not explicitly involve spatial degree of freedom, which in fact implies that any two pairs of interacting particles with the same quark content have the same spatial configuration in despite of hadron environments, such as the $Q\bar{Q}$ in the conventional mesons and multiquark states. The dynamical calculations on the heavy-mesons and the states $[Q_1Q_2][\bar{Q}_3\bar{Q}_4]$ indicate that the difference of their distances are apparent, see Tables~\ref{spectra} and \ref{rms}, which is contradict with the CMIM assumption of the same spatial configuration. Furthermore, it can be found from Table~\ref{4Q} and \ref{part} that the color-magnetic interactions of the states $[Q_1Q_2][\bar{Q}_3\bar{Q}_4]$ in the CMIM are overestimated relative to that in the dynamical models due to the spatial assumption, which results in the appearance of the bound states in the CMIM~\cite{cmwj}. In addition, the difference of the confinement potential based on string and junction $\Delta\langle V^{con}_{min}(4)\rangle$ is not a constant, which depends on the specific state. However, the added constant term $S$ may be thought of as representing the contribution of two additional QCD strings and one junction~\cite{karliner}. In this way, the predictive power of the color-magnetic mechanism needs to be checked on a large scale by more sophisticated models with various QCD dynamic effects.

The ground states $[Q_1Q_2][\bar{Q}_3\bar{Q}_4]$ prefer the color configuration $\bar{\mathbf{3}}_c\otimes\mathbf{3}_c$ in the color-magnetic mechanism \cite{berezhnoy,karliner}. However, the interactions between the $[Q_1Q_2]_{\mathbf{6}_c}$ and $[\bar{Q}_3\bar{Q}_4]_{\bar{\mathbf{6}}}$ in the color configuration $\mathbf{6}_c\otimes\bar{\mathbf{6}}_c$ are attractive although the interactions in the $[Q_1Q_2]_{\mathbf{6}_c}$ ($[\bar{Q}_3\bar{Q}_4]_{\mathbf{6}}$) are repulsive. The attractive interactions are much stronger than those of the $\bar{\mathbf{3}}_c\otimes\mathbf{3}_c$ because the strength of the interaction depends on the color factors listed in Table~\ref{color-matrix}. Therefore, the final result, which is mainly dominated by the Coulomb interaction, of the $\mathbf{6}_c\otimes\bar{\mathbf{6}}_c$ relies on the distance $\langle\mathbf{X}^2\rangle^{\frac{1}{2}}$ between the $[Q_1Q_2]_{\mathbf{6}_c}$ and $[\bar{Q}_3\bar{Q}_4]_{\bar{\mathbf{6}}}$. The heavier the heavy quark mass, the smaller the distance $\langle\mathbf{X}^2\rangle^{\frac{1}{2}}$, the stronger the Coulomb interaction, the bigger the proportion of the $\mathbf{6}_c\otimes\bar{\mathbf{6}}_c$, which can be found from the group $[cc][\bar{c}\bar{c}]$-$[cc][\bar{b}\bar{b}]$-$[bb][\bar{b}\bar{b}]$ with $0^+$ in Tables \ref{part} and
\ref{rms}.

The $\bar{\mathbf{3}}_c\otimes\mathbf{3}_c$ and $\mathbf{6}_c\otimes\bar{\mathbf{6}}_c$ can couple each other through mainly the color-magnetic interaction, the strength of which is inversely proportional to the interacting quark masses. The proportion of the $\mathbf{6}_c\otimes\bar{\mathbf{6}}_c$ in the CQM is bigger than that in the MCFTM because the confinement potential involving the color factor $\langle\mathbf{\lambda}^c_i\cdot\mathbf{\lambda}^c_j\rangle$ in the CQM can strengthen the coupling in the two color configurations, see Table~\ref{4Q}. In the CMIM, the proportion in the group  $[cc][\bar{c}\bar{c}]$-$[cc][\bar{b}\bar{b}]$-$[bb][\bar{b}\bar{b}]$ with $0^+$ in Tables \ref{part} does not change because it only determined by spin-color structure due to the absence of the spatial degree of freedom. In a word, the color configuration $\mathbf{6}_c\otimes\bar{\mathbf{6}}_c$ can not be ignored but should be taken seriously in the investigation on the ground fully-heavy tetraquark states, which is supported by the conclusions of other two models with QCD dynamical effects~\cite{gjwang}.

The $[Q_1Q_2]$ and $[\bar{Q}_3\bar{Q}_4]$ are both considered as a compound object with no internal orbital excitations, namely $l_a=l_b=0$. In the case of the excited states $[Q_1Q_2][\bar{Q}_3\bar{Q}_4]$, the orbital angular excitations are assumed to occur only between the $[Q_1Q_2]$ and $[\bar{Q}_3\bar{Q}_4]$ in the present numerical calculations. Therefore, the total orbital angular momentum of the states $[Q_1Q_2][\bar{Q}_3\bar{Q}_4]$ $L$ is equal to $l_{ab}$ and the parity is $P=(-1)^L$. In Table \ref{proportion}, we present the numerical results of the states $[cc][\bar{c}\bar{c}]$, $[cc][\bar{b}\bar{b}]$ and $[bb][\bar{b}\bar{b}]$ with $L=0$, 1 and 2 and $S=0$. It can be found from Tables \ref{rms} and \ref{proportion} that the sizes of the $[Q_1Q_2]$ and $[\bar{Q}_3\bar{Q}_4]$, $\langle\mathbf{r}_{12}^2\rangle^{\frac{1}{2}}$ and
$\langle\mathbf{r}_{34}^2\rangle^{\frac{1}{2}}$, do not dramatically change with the increase of $L$ and $S$. However, the distance between the $[Q_1Q_2]$ and $[\bar{Q}_3\bar{Q}_4]$ $\langle\mathbf{X}^2\rangle^{\frac{1}{2}}$ rapidly increase with the increase of $L$ in the excited states. It can be concluded form the average distances that the spatial configuration of the states $[Q_1Q_2][\bar{Q}_3\bar{Q}_4]$ is a compact three-dimension structure. In the ground states $(L=0)$, the shape looks like an ellipsoid because the $[Q_1Q_2]$ and $[\bar{Q}_3\bar{Q}_4]$ overlap very strongly each other. In the excited states, the ellipsoid gradually expands to be a dumbbell-like shape with the increase of the distance $\langle\mathbf{X}^2\rangle^{\frac{1}{2}}$ because of the increase of $L$.

The Coulomb interaction between the $[Q_1Q_2]$ and $[\bar{Q}_3\bar{Q}_4]$ rapidly decrease with the increase of the distance $\langle\mathbf{X}^2\rangle^{\frac{1}{2}}$. The $\mathbf{6}_c\otimes\bar{\mathbf{6}}_c$ decreases faster than the $\bar{\mathbf{3}}_c\otimes\mathbf{3}_c$ because of the bigger interaction strength between the $[Q_1Q_2]_{\mathbf{6}_c}$ and $[\bar{Q}_3\bar{Q}_4]_{\bar{\mathbf{6}}}$. The Coulomb interaction in the $\bar{\mathbf{3}}_c\otimes\mathbf{3}_c$ is stronger than that in the $\mathbf{6}_c\otimes\bar{\mathbf{6}}_c$ because the Coulomb interaction in the $[Q_1Q_2]_{\bar{\mathbf{3}}}$ and $[\bar{Q}_3\bar{Q}_4]_{\mathbf{3}}$ is strong attractive while that in the $[Q_1Q_2]_{\mathbf{6}_c}$ and $[\bar{Q}_3\bar{Q}_4]_{\bar{\mathbf{6}}}$ is repulsive. In addition, the kinetic $E_k$ of the $\bar{\mathbf{3}}_c\otimes\mathbf{3}_c$ is obvious lower, more than 100 MeV, than that of the  $\mathbf{6}_c\otimes\bar{\mathbf{6}}_c$ because of the big distance $\langle\mathbf{X}^2\rangle^{\frac{1}{2}}$ induced by the relative weak Coulomb interaction between the $[Q_1Q_2]_{\bar{\mathbf{3}}}$ and $[\bar{Q}_3\bar{Q}_4]_{\mathbf{3}}$, see Table~\ref{proportion}, which is the main reason resulting in the mass difference between the $\bar{\mathbf{3}}_c\otimes\mathbf{3}_c$ and $\mathbf{6}_c\otimes\bar{\mathbf{6}}_c$. The coupling effect between the $\bar{\mathbf{3}}_c\otimes\mathbf{3}_c$ and $\mathbf{6}_c\otimes\bar{\mathbf{6}}_c$ is very weak because of the weak color-magnetic interaction in the excited states. In this way, the proportion of the $\mathbf{6}_c\otimes\bar{\mathbf{6}}_c$ is small in the excited states while the $\bar{\mathbf{3}}_c\otimes\mathbf{3}_c$ is absolutely dominant.

Other different versions of non-relativistic quark models involving the OGE interaction and various type of color confinement potential were also employed to investigate the fully-heavy tetraquark states~\cite{lloyd,gjwang,xhzhong}, which presented similar mass spectra to our models. The masses of the ground states in those quark models are much higher, about 300-500 MeV, than the corresponding thresholds, which indicates that there does not exist a bound state in the scheme of those quark models. However, the non-relativistic model with a Cornell-inspired potential, in which a four-body problem is simplified into three two-body problems, predicted that the lowest S-wave $[cc][\bar{c}\bar{c}]$ might be below their thresholds of spontaneous dissociation into low-lying charmonium pairs~\cite{cpc}.

\subsection{$J/\Psi$-pair resonances observed by LHCb and the fully-heavy state $[cc][\bar{c}\bar{c}]$}

The central value of the broad structure ranging from 6.2 to 6.8 GeV locates at around 6490 MeV~\cite{exp4c2}. The mass and width of the structure $X(6900)$ are
\begin{eqnarray}
M=6905\pm11\pm7~\mbox{MeV},~\Gamma=80\pm19\pm33~\mbox{MeV} \nonumber
\end{eqnarray}
assuming no interference with the nonresonant single-parton scattering continuum~\cite{exp4c2}. The mass and width are changed to
\begin{eqnarray}
M=6886\pm11\pm11~\mbox{MeV},~\Gamma=168\pm33\pm69~\mbox{MeV} \nonumber
\end{eqnarray}
when assuming the nonresonant single-parton scattering continuum interferes with the broad structure~\cite{exp4c2}. The two structures are  obviously higher than the CMIM predictions on the state $[cc][\bar{c}\bar{c}]$, see Table~\ref{4Q}. In this way, the two structures are difficult to be accommodated in the CMIM.

The mass of the ground state $[cc][\bar{c}\bar{c}]$ ranges from 6370 MeV to 6600 MeV in the MCFTM, CQM and other non-relativistic quark models with various dynamical effects~\cite{lloyd,gjwang,xhzhong,segovia}, see Table \ref{ccccground}, which is supported by QCD sum rule~\cite{qcdsum4c}. Matching the central value of the broad structure ranging from 6.2 to 6.8 GeV with the model results, the structure can be described as the ground state $[cc][\bar{c}\bar{c}]$ in the models. However, the models give different quantum numbers. In the MCFTM and the model II in the literature~\cite{gjwang}, the ground states $[cc][\bar{c}\bar{c}]$ with $1^+$ and $2^+$ are both close to the central value. The masses of the ground state $[cc][\bar{c}\bar{c}]$ with $0^+$, $1^+$ and $2^+$ are very close each other, around 6500 MeV, in the model in the literature~\cite{xhzhong}. In the CQM, the ground state $[cc][\bar{c}\bar{c}]$ with $0^+$ has a mass of 6491 MeV, which is very consistent with the value. In the model in the literature~\cite{segovia}, the mass of the ground tetraquark state $[cc][\bar{c}\bar{c}]$ with $2^+$ is closest to the central value. The matching inconsistence results from the slight difference of the model dynamics.
\begin{table}[ht]
\caption{Masses of the ground state $[cc][\bar{c}\bar{c}]$ in various models, unit in MeV.} \label{ccccground}
\begin{tabular}{ccccccccccc}
\toprule[0.8pt] \noalign{\smallskip}
$J^P$&MCFTM&CQM&\cite{lloyd}&I,~II~\cite{gjwang}&\cite{xhzhong}& \cite{segovia} &\cite{qcdsum4c}  \\
\noalign{\smallskip}
\toprule[0.8pt]
\noalign{\smallskip}
$0^+$&6407&6491&6477&6377,~6371&6470&6350&$6440\pm0.15$\\
 \noalign{\smallskip}
$1^+$&6463&6580&6528&6425,~6450&6512&6440&$6370\pm0.18$\\
 \noalign{\smallskip}
$2^+$&6486&6607&6573&6432,~6479&6534&6470&$6370\pm0.19$\\
 \noalign{\smallskip}
\toprule[0.8pt]
\end{tabular}
%\end{table}
%\begin{table}[ht]
\caption{Masses of the excited state $[cc][\bar{c}\bar{c}]$ in the CQM and MCFTM, unit in MeV.} \label{ccccexcited}
\begin{tabular}{ccccccccccc}
\toprule[0.8pt] \noalign{\smallskip}
$L$&&1&&&2&& \\
\noalign{\smallskip}
$S$&0&1&2&0&1&2  \\
\noalign{\smallskip}
\toprule[0.8pt]
\noalign{\smallskip}
CQM&6901&6912&6924&7182&7185&7191\\
\noalign{\smallskip}
MCFTM&~~~~6727~~~~&6735&~~~~6744~~~~&6944&~~~~6947~~~~&6951 \\
\noalign{\smallskip}
\toprule[0.8pt]
\end{tabular}
\end{table}

The masses of the excited state $[cc][\bar{c}\bar{c}]$ with $L=1$, 2 and $S=0$, 1, 2 are calculated in the MCFTM and CQM, which are given in Table \ref{ccccexcited}. The spin-orbital interaction is not taken into account in the present calculation because it is very weak, about several MeV~\cite{cscs}. One can find that the effect of the spin-dependent interaction on the masses of the excited state with different total spin is small. The masses in the CQM are apparent higher than those in the MCFTM, about
175 MeV for the excited states with $L=1$ and about 240 MeV for the excited states with $L=2$, which originates from the different forms of confinement potential. In the CQM, the masses of the excited state $[cc][\bar{c}\bar{c}]$ with $L=1$ are completely consistent with that of the narrow structure $X(6900)$. Therefore, the structure can be described as the excited state $[cc][\bar{c}\bar{c}]$ with $L=1$ in the CQM, which is supported by the NRPQM~\cite{p-wave}. However, the masses of the excited states $[cc][\bar{c}\bar{c}]$ with $L=1$ in the MCFTM are obvious lower than that of the narrow structure while the excited states $[cc][\bar{c}\bar{c}]$ with $L=2$ are very close to the narrow structure.

%QCD sum rule study suggested that the broad structure around 6.2-6.8 GeV can be interpreted as an $S$-wave $cc\bar{c}\bar{c}$ tetraquark state with $J^{PC}=0^{++}$ or $2^{++}$, and %the $X(6900)$ can be interpreted as a $P$-wave one with $J^{PC}=0^{-+}$ or $1^{-+}$~\cite{x6900sumrule}.

\section{summary}

In this work, we use the CMIM and two quark models with the OGE interaction and color confinement potential, CQM with two-body confinement and MCFTM with multibody one based on
the lattice color flux-tube picture, to systematically investigate the properties of the states $[Q_1Q_2][\bar{Q}_3\bar{Q}_4]$. The difference between the two confinement potentials in the ground states is 30-120 MeV. In the excited states with $L\le2$, the difference is around 200 MeV.% The multibody confinement potential is usually thought to be closer to real physical images than two-body one related to color charges.

The masses of the ground states $[Q_1Q_2][\bar{Q}_3\bar{Q}_4]$ predicted by the CMIM are close to the corresponding two heavy-meson threshold, which is much lower, about hundreds
of MeV, than the masses predicted by the models with QCD dynamic effects mainly because of the strong Coulomb interaction. Therefore, the CMIM can not completely absorb QCD dynamic effects. In addition, the CMIM may overestimate the color-magnetic interaction in the extension from heavy mesons to the states $[Q_1Q_2][\bar{Q}_3\bar{Q}_4]$ because of the assumption of the same spatial configurations. Therefore, the reliability of the CMIM extension from the conventional hadrons to multiquark states needs further study.

The Coulomb interaction plays an important role in the dynamical model calculation on the heavy mesons. The interaction in the states $[Q_1Q_2][\bar{Q}_3\bar{Q}_4]$ is weaker than that of the corresponding threshold of two heavy meons, which directly induces the fact that there does not exist a bound state $[Q_1Q_2][\bar{Q}_3\bar{Q}_4]$ in the dynamical models. The color configuration $\left[[Q_1Q_2]_{\mathbf{6}_c}[\bar{Q}_3\bar{Q}_4]_{\bar{\mathbf{6}}_c}\right]_{\mathbf{1}}$ can not be ignored in the ground states owing to the strong Coulomb interaction between the $[Q_1Q_2]_{\mathbf{6}_c}$ and $[\bar{Q}_3\bar{Q}_4]_{\bar{\mathbf{6}}_c}$ at the short distance. The color configuration $\left[[Q_1Q_2]_{\bar{\mathbf{3}}_c}[\bar{Q}_2\bar{Q}_4]_{\mathbf{3}_c}\right]_{\mathbf{1}}$ is absolutely dominant in the excited states because the Coulomb interaction in the $[Q_1Q_2]_{\bar{\mathbf{3}}}$ and $[\bar{Q}_3\bar{Q}_4]_{\mathbf{3}}$ is strong attractive while that in the $[Q_1Q_2]_{\mathbf{6}_c}$ and $[\bar{Q}_3\bar{Q}_4]_{\bar{\mathbf{6}}}$ is repulsive.

The $J/\Psi$-pair resonances observed by LHCb are difficult to be accommodated in the CMIM because their masses are much higher than the CMIM predictions on the state $[cc][\bar{c}\bar{c}]$. The broad structure ranging from 6.2 to 6.8 GeV can be described as the ground tetraquark state $[cc][\bar{c}\bar{c}]$ in the various quark models.
However, it has different quantum numbers because of the slight difference of the dynamics in the models. The narrow structure $X(6900)$ can be described as the excited state $[cc][\bar{c}\bar{c}]$ with $L=1$ in the CQM. However, the masses of the states $[cc][\bar{c}\bar{c}]$ with $L=2$ in the MCFTM are very close to that of the narrow structure. Although many theoretical investigations have been devoted to those structures, the interpretations on the natures of the two structure are not completely consistent in the different theoretical frames so far. Therefore, more data along with additional measurements, including determination of the spin-parity quantum numbers, are needed, which will contribute immeasurably to an understanding of the properties of these structures.

\acknowledgments
{This research is partly supported by the National Science Foundation of China under Contracts Nos. 11875226 and 11775118, the Chongqing Natural Science Foundation under Project No. cstc2019jcyj-msxmX0409 and Fundamental Research Funds for the Central Universities under Contracts No. SWU118111.}

\end{document}